\documentclass[aps,twocolumn,groupedaddress,prl]{revtex4}
\usepackage{amsfonts}
\usepackage{amsmath}
\usepackage{amssymb}
\usepackage{graphicx}

\begin{document}

\title{Chern number spins of Mn acceptor magnets in GaAs}
\author{T.O. Strandberg$^{1}$}

\author{C.M. Canali$^{1}$}%


\author{A.H. MacDonald$^{2}$}

\affiliation{%
$^{1}$School of Computer Science, Physics and Mathematics, Linn\ae us University. Sm\.{a}landsgatan 26 39233 Kalmar-Sweden.}%

\affiliation{$^{2}$Department of Physics, University of Texas at Austin, Austin, Texas 78712, USA
}%

\date{\today}

\begin{abstract}
We determine the effective total spin $J$ of local moments 
formed from acceptor states bound to Mn ions in GaAs by evaluating their magnetic Chern numbers.
We find that when individual Mn atoms are close to the sample surface, the 
total spin changes from $J = 1$ to $J = 2$, due to quenching of the acceptor orbital moment. 
For Mn pairs in bulk, the total $J$ depends on the pair orientation 
in the GaAs lattice and on the separation between the Mn atoms. 
We point out that Berry curvature variation
as a function of local moment orientation can
profoundly influence the quantum spin dynamics of these magnetic entities. 

\end{abstract}

\pacs{Valid PACS appear here}
\maketitle


State-of-the-art STM techniques have made it possible to substitute transition metal impurities 
for individual atoms in semiconductor crystals. 
\cite{kitchen_jsc05, yazdani_nat06}
When the impurity behaves as a dopant, high-resolution STM scanning 
capabilities can then provide detailed information
on the nature of the bound donor or 
acceptor 
states.\cite{yakunin_prl04, kitchen_jsc05, yakunin_prl05, yazdani_nat06, 
wiesendanger_MnInAs, koenraad_prb08, garleff_prb_2010} 
Advances in spin-polarized STM techniques are now making it possible to address the quantum spin dynamics 
of individual coupled acceptor-impurity systems.\cite{wiesend_RMP_09}   
These centers represent a new class of magnetic entities which we refer to as 
{\it donor magnets} or {\it acceptor magnets}.  They have precisely reproducible properties
that are intermediate in character between those of  
atomic local moments and nanomagnets, have promise for applications in spintronics
and quantum information processing, and act as the building blocks of
ferromagnetism in 
semiconductors\cite{jungw_rmp06, uppsala_DMSreview_2010, zunger_physics_2010} 
that are doped with many transition metal impurities.

The interpretation of present and future experiments requires a theoretical understanding 
of the quantum dynamics of acceptor and donor magnets. 
For the specific case of individual Mn impurities in bulk
GaAs it is known that the ground-state total angular momentum of the Mn centers is J=1. 
This value is the result of antiferromagnetic
coupling between the localized S=5/2 Mn spin and the spin (s= 1/2) and orbital moment (l=1) of the acceptor hole.  
The $J=1$ character of the ground-state effective spin of the Mn embedded atom is 
supported by ESR and infrared spectroscopy experiments.\cite{schneider_prb87, linnarsson_prb97}
Several questions concerning the magnetic properties of Mn dopants in GaAs nevertheless
remain unanswered: 
(i) What happens
to the total angular momentum $J$ when the dopant is close to the symmetry breaking surface which 
provides STM access? (ii) Is there an effective "giant spin" describing 
the low-energy magnetic properties of two or more nearby embedded Mn impurities in GaAs, and in that case
what is its value (iii) Can we determine an effective spin Hamiltonian
describing the quantum dynamics of acceptor magnets? 
The answers to these questions depend on a complex interplay between
the kinetic exchange that couples Mn and acceptor spins, and the variation of the 
acceptor-level orbital spinor with Mn spin orientation 
which is controlled by spin-orbit interactions (SOI)
and the crystalline environment.  

In this Letter we present a possible answer to these questions.
The approach we use is  
similar in spirit to ones used to quantize the slow vibronic degrees
of freedom in molecular systems.\cite{geometricquantumphases}
It can be used to quantize magnetization dynamics 
in any theory in which the magnetization direction is initially treated 
as a classical parameter, for example spin-density-functional theory. 
It identifies the 
effective total spin $J$ of the Mn acceptor magnet with a topological Chern number
which is the average of a Berry curvature electronic functional 
over all possible directions of the Mn acceptor magnetic moment.
The procedure yields the expected 
$J=1$ value for Mn in bulk GaAs. However when a Mn atom is close to a symmetry-breaking surface,
the circumstance most commonly studied in STM experiments, 
we find that $J=2$ because the orbital contribution of the acceptor is quenched. Mn pairs close to a surface always have $J=4$,
due to strong localization of the acceptor wave-function. Surprisingly, for Mn pairs in bulk GaAs 
we find that the total spin can switch between $J=4$, $J=3$, and $J=2$ depending on the orientation of the pair in the crystal 
and the distance between the two Mn atoms. Our theory allows us to extract a quantum spin Hamiltonian for the magnetic
centers. The spectrum of these Hamiltonians can be strongly affected by Berry curvature variation as a 
function of magnetization orientation, which is especially strong whenever there is a weakly avoided
level crossing at the Fermi energy.

\begin{figure}[ptb]
\resizebox{6cm}{!}{\includegraphics{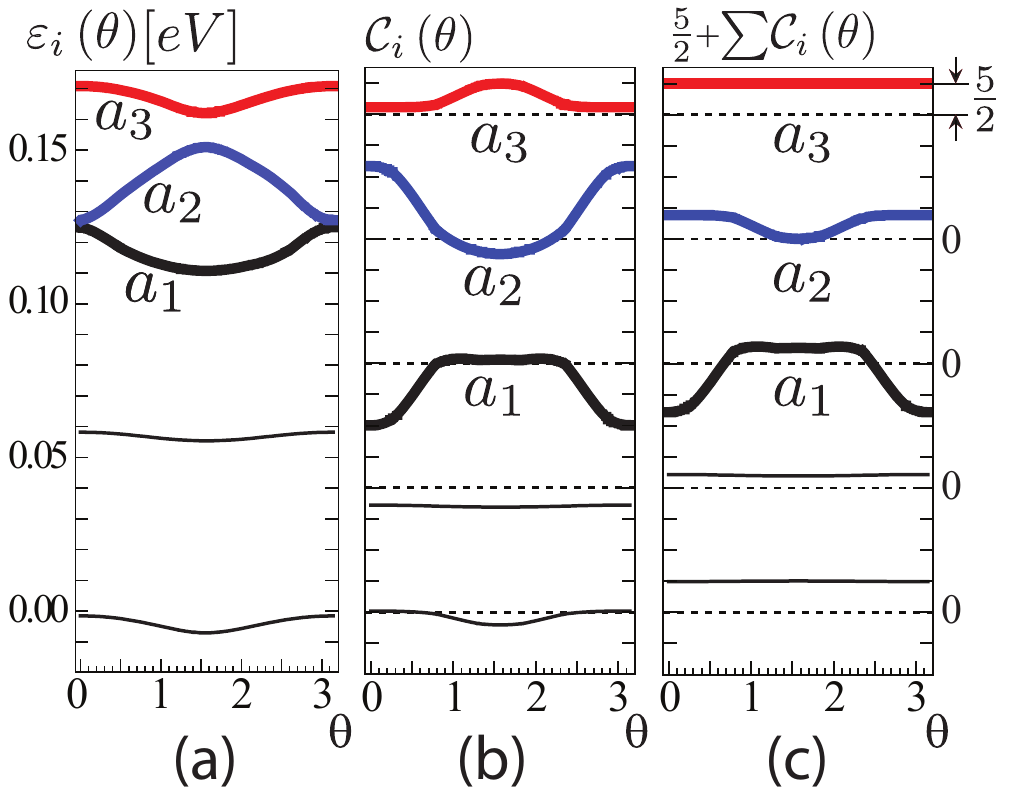}}
\caption{(Color online)\
Electronic structure and Berry curvature for one Mn impurity in bulk GaAs as a function of the polar angle $\theta$
specifying the Mn magnetic moment
direction relative to a cubic crystal axis.
(a) The three acceptor energy levels $a_1$, $a_2$, $a_3$ above the valence band edge.
The topmost level $a_3$ (red curve) is the
only one occupied by a hole; $a_2$ (blue curve) is the highest energy occupied orbital. 
(b) Berry curvatures for the individual levels in (a).
The dashed lines mark the zero for a given curvature plot.  
(c) Cumulative level curvature plus the constant contribution 5/2 from the Mn impurity.
The scale for curvature plots is indicated on the right axis of panel (c).
The blue curve, in which the $a_3$ contribution is not included, is the one that is 
relevant to the Mn acceptor magnet.
Adding the curvature of the acceptor level $a_3$ to the total curvature gives a constant equal to $5/2$ (flat red curve).}
\label{fig1}%
\end{figure}

We start by introducing a microscopic tight-binding model that captures the salient electronic properties of Mn impurities
in GaAs.\cite{scm_MnGaAs_paper1_prb09, scm_MnGaAs_paper2_prb2010} The Hamiltonian reads
\begin{equation}
\label{ham}
H = H_{\rm band} + H_{\rm SO} + J_{pd}\sum_{m}\sum_{n[m]}\vec{s}_{n}\cdot \vec{S}_{m}
\end{equation}
where $H_{\rm band}$ contains the Slater-Koster parameters that reproduce the band structure of bulk GaAs
plus parameters that account for the $4s$ and $4p$ orbitals of each substitutional Mn atom; the second
term is a one-body atomic spin-orbit term. The third term describes an effective antiferromagnetic
exchange interaction between a Mn spin $\vec S_m$ and its nearest-neighbor As 
$p$-spins ${\vec s}_n $. This is a kinetic $pd$ exchange originating from the hybridization of the Mn 3d-levels with the
As p-levels.  
We also include a spin-independent Coulombic potential to represent the Coulomb potential associated with 
the Mn ion\cite{tangflatte_prl04,tangflatte_prb05,scm_MnGaAs_paper1_prb09};  we do not explicitly include the Mn $3d$ orbitals,  
and $\vec S_m$ represents a {\it classical} vector of magnitude $5/2$.

The Coulomb potential and level repulsion due to hybridization with the Mn d-levels together 
push acceptor states whose spins are aligned with the Mn magnetic moments
above the valence-band edge.\cite{tangflatte_prl04,tangflatte_prb05,scm_MnGaAs_paper1_prb09}
Each Mn impurity introduces three acceptor levels ($p_x$, $p_y$, $p_z$) 
which would be degenerate in the absence of SOI.
SOI's not only lift the degeneracy, but also 
lead to a dependence of energies and orbitals 
on the direction of the Mn magnetic moment.
For a
neutral Mn impurity only the top most of these three states is occupied by a hole. 
Fig.~\ref{fig1}(a) shows that this
electronic structure is reproduced by the model of Eq.~\ref{ham} implemented numerically for one substitutional Mn impurity
in the middle of a 1200-atom GaAs cluster with periodic boundary conditions.  
For a pair of Mn atoms, the lowest energy state of the system is 
usually the one in which the two Mn magnetic moments
are ferromagnetically aligned.\cite{scm_MnGaAs_paper2_prb2010}
For this configuration, the two sets of acceptor states
form bonding and antibonding molecular orbitals.
The two topmost empty levels are then 
split by an energy that is related to effective Mn-Mn exchange interaction, 
and varies strongly
with the pair orientation and the distance between the two Mn.\cite{yazdani_nat06, scm_MnGaAs_paper2_prb2010}

This electronic structure sets the stage for the central part of our study in which we address
the quantum spin dynamics of the Mn centers. 
The starting point to derive an effective ``total spin'' Hamiltonian 
is an approximate imaginary-time quantum action for the coherent spin magnetization direction 
$\hat n$\cite{carlo_prl03,scm_prb08}
\begin{equation}
S[\hat{n}]\equiv\int d\tau\left[  \langle\Psi\lbrack\hat{n}]|\mathbf{\nabla
}_{\hat{n}}\Psi\lbrack\hat{n}]\rangle\cdot\frac{\partial\hat{n}}{\partial\tau
}+E[\hat{n}]\right]\,  .
\label{action}%
\end{equation}
In Eq.~(\ref{action}),
$|\Psi\lbrack\hat{n}]\rangle$ is the many-particle ground-state wave function
obtained by diagonalizing the Hamiltonian (\ref{ham}), and $E[\hat{n}]$ is 
the total energy obtained by summing over occupied single-particle states. 
Here $\hat n$  represents the orientation of the Mn spin vectors $\vec S_m$,
which are assumed to be ferromagnetically aligned when 
the Mn ions are not isolated.
The first term in (\ref{action}) is a Berry
phase term, which for a closed path $\gamma$ on the unit sphere is given by %
$\mathcal{P}=i%
{\displaystyle\oint\nolimits_{\gamma}}
d\hat{n}\cdot\langle\Psi|\mathbf{\nabla}_{\hat{n}}\Psi\rangle $.
The line integral can be converted into an
integral over enclosed area with a gauge-invariant integrand 
known as the Berry
curvature\cite{resta2000,auerbach94:_inter_elect_quant_magnet},
\begin{equation}
\mathcal{\vec{C}}[\hat{n}]=i\mathbf{\nabla}_{\hat{n}}\times\langle
\Psi|\mathbf{\nabla}_{\hat{n}}\Psi\rangle.
\end{equation}
In our model, $\mathcal{\vec{C}}[\hat{n}] = \sum_i^{\rm occ} \mathcal{\vec{C}}_i[\hat{n}]$, the sum of the 
Berry curvatures $\mathcal{\vec{C}}_i[\hat{n}]$ of all occupied single-particle 
levels $i= 1,2, \dots, {\rm occ}$. 
In the absence of SOI, all  $\mathcal{C}_i = \mathcal{\vec{C}}_i[\hat{n}]\cdot \hat n$
are constant and equal to the spin projection of each level, $\pm 1/2$.
In this case $\mathcal{\vec{C}}\cdot \hat n = \sum_i^{\rm occ} \pm \frac{1}{2}= S$, where $S$ is the total spin of the system.
The effect of SOI is twofold.
First, there is now an orbital contribution to the curvature. Second, $\mathcal{\vec{C}}_i[\hat{n}]$ 
varies with $\hat n$, as shown in Fig.~\ref{fig1}(b). 

The average of the Berry curvature over the unit sphere of all possible directions,
\begin{equation}
J =\tfrac{1}{4\pi}\int_{S^{2}}\mathcal{\vec{C}}[\hat{n}]\cdot\hat,
{n}\;dA\; ,
\end{equation}
is a topological invariant, known as first Chern number.\cite{simon1983} Although $\mathcal{\vec{C}}[\hat{n}]$
can fluctuate strongly, $J$ is always a half-integer; its value can change only if the system
suffers a level crossing at the Fermi level.\cite{ carlo_prl03,scm_prb08} 
It is useful to introduce single-level Chern numbers
$j_i=\tfrac{1}{4\pi}\int_{S^{2}}\mathcal{\vec{C}}_i[\hat{n}]\cdot\hat
{n}\;dA$,
so that $J = \sum_i ^{\rm occ}j_i$. When SOI's are included $j_i$ can be different from $\pm 1/2$.
The contribution from the spin-polarized Mn d-orbitals that are not explicitly included in our 
calculation are accounted for by adding $5/2$ to the total Berry curvature for each Mn ion in the system.
As we explain below,
the total Chern number $J$ plays the role of the effective quantum spin of the Mn acceptor magnetic.

We first discuss the case of one Mn impurity in GaAs and confirm 
that our approach yields sensible results. 
Indeed, as shown in Table I, 
when the Mn is located in the bulk, solving our model numerically for a 1200-atom cluster
gives $J = 1$, as expected.  To understand this result, we observe that the total Chern number for a 
full valence band vanishes.  The total Chern number of the acceptor magnet is therefore 
$J = 5/2 - j$, where $j$ is the Chern number of the topmost acceptor state which is occupied
by a hole. As shown in Table I, we find $j = 3/2 = 1/2 +1$.  The orbital contribution to $j$, which is absent 
in any theory that does not include spin-orbit interactions appears in our approach because the orbital 
content of the topmost acceptor state varies with moment orientation.  The orbital moment is locked to the 
spin-moment by strong SOI's.  The result is different however, when the Mn impurity is located on the (110) 
surface,\cite{note1}
a case frequently considered in STM experiments.
For this case we find that the orbital content of the acceptor wavefunction 
does not vary substantially with 
magnetization direction. The Chern number of the acceptor
level only has a $j= 1/2$ spin contribution so that $J = 5/2 - 1/2 = 2$. 
Evidently symmetry breaking on the (110) surface, with only three nearest-neighbor As atoms instead of four, creates
a local environment for a Mn whose symmetry is lower than the tetragonal symmetry seen by the impurity
in bulk GaAs. As a result the acceptor
orbital moment is completely quenched.
(A direct calculation shows that $\langle a_3 |{\vec L}|a_3\rangle =0$.) Interestingly, our calculations show that $j$ remains equal to $1/2$ also 
when the Mn is located on one of the immediate sub-surfaces below the top (110) surface. 
We expect that $j$ should eventually switch to the bulk value when the Mn is located deeply below the surface,
but this does not happen for the film thicknesses in the present simulation.  

We can now investigate magnetic clusters with two Mn atoms, where the resulting total spin is less intuitive.  
\begin{figure}[ptb]
\includegraphics[height = 4.5 cm, width = 7 cm]{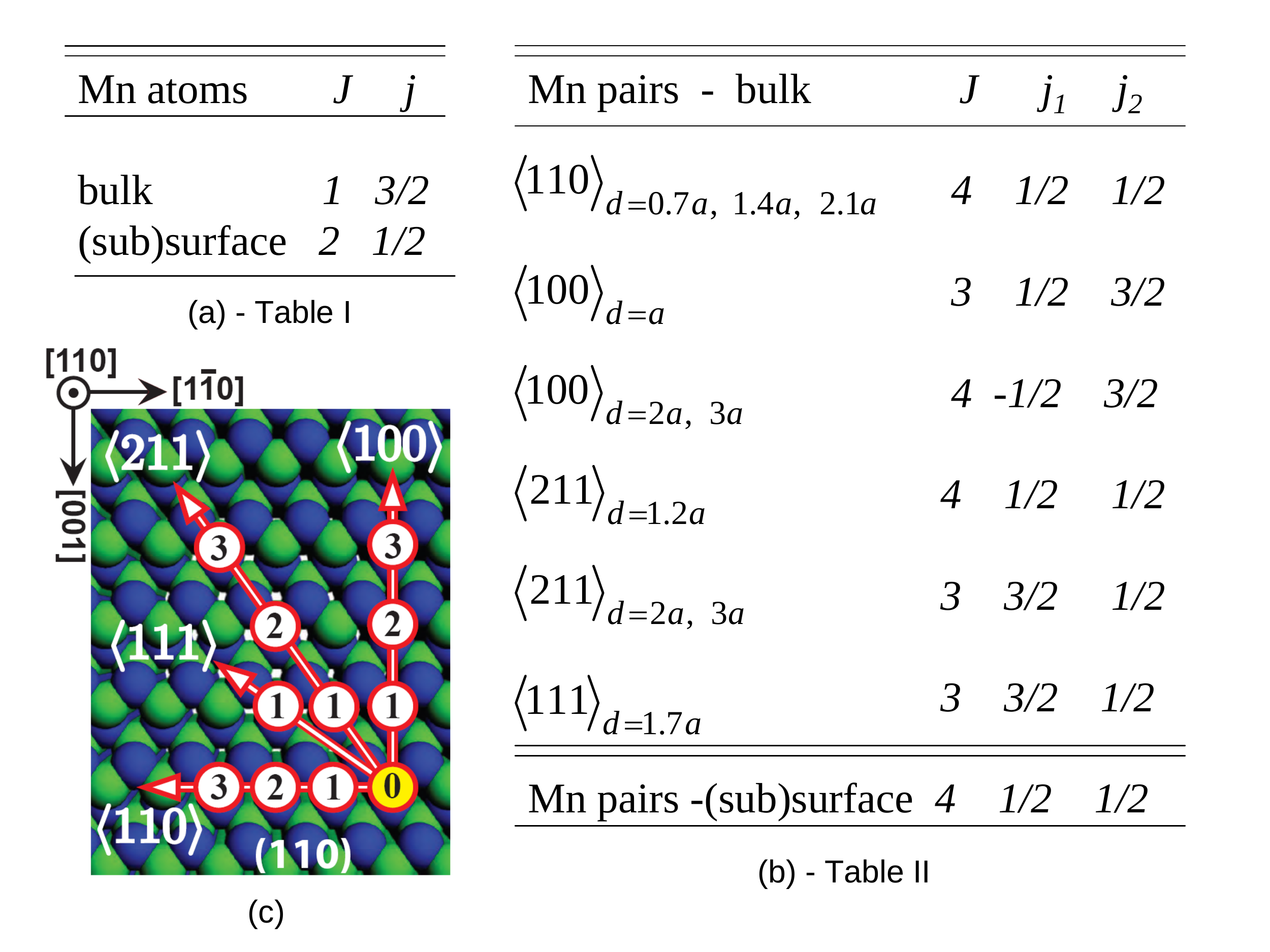}
\caption{(Color online)\  Calculated Chern numbers for Mn impurities in bulk GaAs or in the (110) surface and nearby sub-surfaces.
(a) Results for one Mn impurity. $J$ is the total Chern number and $j$ the Chern number of the acceptor level.
(b) Results for various Mn pairs $\langle lmn\rangle_d$ separated by distance 
$d$ in units of the lattice constant $a$. 
$j_{1}$ and $j_{2}$ are the two acceptor Chern numbers. (c) 
Orientation $\langle lmn\rangle$  and separation of different Mn pairs on the (110) surface of
GaAs.} 
\label{fig2}%
\end{figure}
Table II 
shows the Chern numbers for several ferromagnetic
Mn pairs, whose orientation $\langle lmn \rangle$ 
in the crystal
and Mn separation $d$ is 
described in Fig.~\ref{fig2}(c). When the pair is positioned on the (110) surface or in one of its nearby sub-layers, we find
that the individual Chern numbers of the two empty acceptor states are always $j_1 = j_2 = 1/2$, yielding 
a total Chern number $J= 2* 5/2 - 2* 1/2 = 4$. The situation for a Mn pair in bulk GaAs is more complex. We can see that, while
for several pairs (e.g. all the $\langle 110\rangle$ pairs) $J=4$, like for the surface case, other pairs have $J=3$, and 
ferromagnetically coupled remote spins should have $J=2$.
In this second case one of the individual acceptor Chern numbers, $j_1$ or $j_2$, is equal to $3/2$, and in the latter case 
both have Chern number $3/2$.  Clearly the presence
of a second Mn affects the orbital magnetic properties of the other, 
in a way that depends both on pair orientation and Mn
separation. The outcome for the pair is not easily predictable. For example,
while the $\langle 100\rangle$ pair switches from $J=3$ at the shortest
Mn separation $d= a$ to $J=4$ at larger separations, the $\langle 211\rangle$ pair behaves exactly 
in the opposite way. ($a$ is the lattice constant.)

In the remaining part of the paper we will extract a quantum spin Hamiltonian 
describing the dynamics associated with the moment orientation of acceptor magnets. 
We return to the action given by Eq.~\ref{action} and perform a change
of variables\cite{carlo_prl03,scm_prb08} 
from $\hat n(\theta, \phi)$ to  $\hat n'(\theta', \phi')$ that transforms the Berry curvature 
$\mathcal{C}[\hat{n}]$  
to a constant $\mathcal{C'}[\hat{n'}]= J$. This change of variables rescales the local
curvature metric such that 
$\mathcal{C}\left(\theta,\phi\right)  \sin(\theta) d\theta d\phi=J \sin(\theta)d\theta^{\prime}d\phi^{\prime
}$. The real-time action for a path becomes
\begin{equation}
\mathcal{S}_{\text{spin}}^{\left(  J\right)  }[\hat{n}^{\prime}]\equiv\int
_{0}^{t}dt^{\prime}\left[  i\;\vec{A}_{J}\cdot\frac{d\hat{n}^{\prime}%
}{dt^{\prime}}-E\{\hat{n}[\hat{n}^{\prime}\left(  t^{\prime}\right)
]\}\right]  , \label{singleaction}%
\end{equation}
where $\vec{A}_{J}=J\hat{\phi}^{\prime}\left(  1-\cos\theta^{\prime}\right)
/\sin\theta^{\prime}.$ Eq.~\ref{singleaction} is the quantum action 
for an effective ``total spin'' quantum number $J$\cite{auerbach94:_inter_elect_quant_magnet}. 
The second term
in the integrand is the semi-classical 
Hamiltonian of the system, which is given by\
\begin{equation}
{\tilde E(\hat n')} = (E\{\hat{n}[\hat{n}^{\prime}  ]\}=\langle J,\hat
{n}^{\prime}  |\tilde {\mathcal{H}}|J,\hat{n}^{\prime}
\rangle
\end{equation}
where $\tilde {\mathcal{H}}$ is the quantum spin Hamiltonian and $|J, {\hat n}'\rangle$ is a spin-$J$ coherent state parametrized
by the unit vector ${\hat n}'(\theta', \phi')$. The function ${\tilde E(\hat n')}$ is
is the anisotropy energy transformed so that it also captures Berry curvature variation.
The quantum Hamiltonian $\tilde {\mathcal{H}}$ is constructed by first expanding  
${\tilde E(\hat n')}$ in spherical harmonics 
${\tilde E(\hat n')}=\sum_{\ell=0}%
^{2J}\sum_{m=-\ell}^{\ell}\gamma_{\ell}^{m}Y_{\ell}^{m}\left(  \hat{n}%
^{\prime}\right)$. We then use a formula\cite{scm_prb08}
which relates the spherical harmonics expansion 
coefficients $\gamma_{\lambda}^{\mu}$ to the matrix elements
${\tilde {\mathcal {H}}}_{mm^{\prime}}^{\left(  J\right)  }=\langle J,m|\tilde {\mathcal{H}}|J,m^{\prime}\rangle$ of
the quantum spin Hamiltonian:
\begin{align}
H_{mm^{\prime}}^{\left(  J\right)  }  &  =\left(  -1\right)  ^{m^{\prime}%
-J}\sum_{\lambda=0}^{2J}\sum_{\mu=-\lambda}^{\lambda}\gamma_{\lambda}^{\mu
}\sqrt{\frac{2\lambda+1}{4\pi}}\nonumber\\
&  \times\left(
\begin{array}
[c]{ccc}%
J & J & \lambda\\
m & -m^{\prime} & \mu
\end{array}
\right)  \left/  \left(
\begin{array}
[c]{ccc}%
J & J & \lambda\\
J & -J & 0
\end{array}
\right)  \right.  \label{quantform}%
\end{align}
where the quantities in parenthesis are Wigner 3J symbols.
Once the Hamiltonian matrix has been obtained, it can be decomposed and rewritten as a
combination of spin operators.\cite{scm_prb08} 

\begin{figure}[ptb]
\resizebox{6cm}{!}{\includegraphics{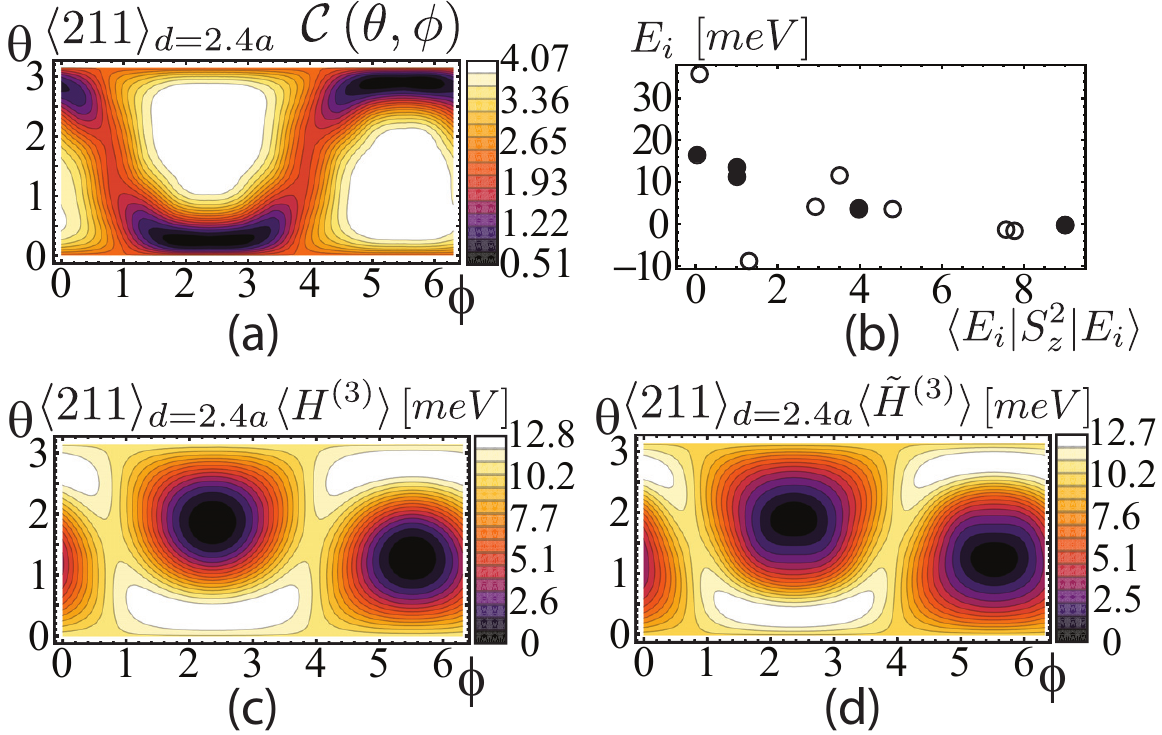}}
\caption{(Color online)\ Effects of the Berry curvature on the quantum spin properties of the 
ferromagnetic Mn pair $\langle 211\rangle_{d=2.4 a}$
in bulk GaAs.
(a) Total Berry curvature landscape as a function of the pair magnetic moment direction.
The two large dips at $\theta \approx 0 ,\pi$, relative to the [001] axis signal avoided crossings at the Fermi level. 
(b) Spectrum of the
effective quantum spin Hamiltonian with Chern number $J= 3$ {\it vs.} the expectation value of $S_z^2$. 
The filled circles
are the spectrum of ${\cal H}^3$ obtained by quantizing the classical magnetic anisotropy landscape.
The  empty circles are the spectrum of ${\tilde {\cal H}}^3$, which includes Berry curvature effects.
(c) and (d) are  the magnetic anisotropy landscape
for the cases when Berry curvature effects are excluded or included respectively.}
\label{fig3}%
\end{figure}

As an example of this procedure, we show in Fig.~\ref{fig3} results for the Mn pair $\langle 211\rangle_{d= 1.4a}$ in
bulk GaAs. 
In panel (a) we plot the Berry curvature functional $\mathcal{C}[\hat {n}] = \mathcal{C}(\theta, \phi)$. 
The coordinate system used for these plots has $\theta=0$ parallel to the [001] axis,
$\left(  \theta=\pi/2,\phi=0\right)  $ parallel to [100], and $\left(
\theta=\pi/2,\phi=\pi/2\right) $ parallel to [010] (see Fig.~\ref{fig2}).
The calculated Chern number
for this pair is $J=3$. The large dips of $\mathcal{C}(\theta, \phi)$ below this value 
for $\theta \approx 0, \pi$  signal the
occurrence of narrowly avoided level crossings at the Fermi level,
for two time-reversed directions. According to our theory, we expect
that in this case Berry phase variations strongly influence the spectrum of the quantum spin Hamiltonian. This is 
indeed the case, as shown in Fig.~\ref{fig3}(b), where we plot the spectrum $\{E_i\}_{i = 1, \dots, 2J+1} $ 
of the Hamiltonian obtained
when Berry phase corrections are either included (${\tilde {\cal H}}^3$, empty circles) 
or absent (${\cal H}^3$, filled circles),
versus 
$\langle E_i|S_z^2| E_i\rangle$.
($S_z$ is the z-component of the effective spin.) 
The difference in the two spectra has implications for the classical magnetic anisotropy landscape.
In Fig~\ref{fig3}(c) and (d) we plot $\langle J, {\vec n}| {\cal H}^3|J, {\vec n}\rangle$  and 
$\langle {J, \vec n}|{\tilde {\cal H}}^3|J, {\vec n}\rangle$ respectively, as a function of  $\theta$ and
$\phi$. 
Although the anisotropy minima are present in both cases, the barrier that
separates them is considerably reduced by Berry phase corrections. The dips in the curvature
at level crossings increase the quantum tunneling rates of the 
magnetization between the two minima. 

In conclusion, we have proposed that the effective total spin of Mn impurities in GaAs is a
topological Chern number that includes both the contribution of the Mn spins and the spin and orbital
moments of the acceptor states.  The effective spin depends sensitively on the environment around the Mn
impurities and it is strongly affected by the presence of symmetry breaking surfaces and the geometry
of the magnetic clusters in the GaAs lattice. The quantum dynamics of the effective spin is 
qualitatively modified by Berry phase corrections caused by electronic degeneracies at some direction of
the Mn magnetic moments. This has implications on the stability of the quasi-classical magnetization of the
magnetic center around minima in the anisotropy energy. 

This work was supported by the Welch Foundation, by the National Science Foundation under grant
DMR-0606489, 
the Faculty of Natural Sciences at Linnaeus University, and by the Swedish
Research Council under Grant No: 621-2007-5019.


%
%


\begin{thebibliography}{1}
\bibitem{note1}  To treat this case, we consider a GaAs cluster with a
(110) surface exposed, and apply periodic boundary conditions
in the other two directions. The surface atoms are
relaxed as described Ref.~\onlinecite{scm_MnGaAs_paper1_prb09}.
\end{thebibliography}

%

\end{document}